\documentclass[letterpaper]{article}
\pdfoutput=1
\usepackage{aaai}
\usepackage{times}
\usepackage{helvet}
\usepackage{courier}
\usepackage{amsmath}
\usepackage{multicol}
\usepackage{multirow}
\usepackage{soul}
\usepackage{enumitem}
\usepackage{xspace}
\usepackage{graphicx}
\usepackage{subfigure}
\usepackage{bm}
\usepackage{algorithm}
\usepackage{algpseudocode}
\usepackage{xcolor}
\usepackage{booktabs}
\usepackage{textcomp}
\usepackage[hyphens]{url}
\usepackage{amsfonts,amssymb}
\usepackage{makecell}

\begin{document}
%
\title{Long Short-Term Preference Modeling for Continuous-Time Sequential Recommendation}
\author{
Huixuan Chi\textsuperscript{\rm 1}\thanks{These authors contributed equally.},
Hao Xu\textsuperscript{\rm 2}\footnotemark[1], 
Hao Fu\textsuperscript{\rm 2}, 
Mengya Liu\textsuperscript{\rm 3}, 
Mengdi Zhang\textsuperscript{\rm 2},
Yuji Yang\textsuperscript{\rm 2},
Qinfen Hao\textsuperscript{\rm 1}\thanks{Corresponding author.},
Wei Wu\textsuperscript{\rm 2} \\
\textsuperscript{\rm 1}{Institute of Computing Technology, Chinese Academy of Sciences, Beijing, China}\\
\textsuperscript{\rm 2}{Meituan Inc, Beijing, China}\\
\textsuperscript{\rm 3}{University of Southampton, Southampton, UK}\\
chihuixuan21s@ict.ac.cn, kingsleyhsu1@gmail.com, haofuch@outlook.com, Mengya.Liu@soton.ac.uk\\
mdzhangmd@gmail.com, yangyujiyyj@gmail.com, haoqinfen@ict.ac.cn, wuwei19850318@gmail.com
}
\maketitle

\newcommand{\modelname}{\textbf{LSTSR}\xspace}
\newcommand{\premodelname}{LSTSR\xspace}
\newcommand{\std}[1]{\textcolor{gray}{\scriptsize{$\pm$#1}}}

\begin{abstract}

Modeling the evolution of user preference is essential in recommender systems. Recently, dynamic graph-based methods have been studied and achieved SOTA for recommendation, majority of which focus on user's stable long-term preference. However, in real-world scenario, user's short-term preference evolves over time dynamically. Although there exists sequential methods that attempt to capture it, how to model the evolution of short-term preference with dynamic graph-based methods has not been well-addressed yet. In particular: 1) existing methods do not explicitly encode and capture the evolution of short-term preference as sequential methods do; 2) simply using last few interactions is not enough for modeling the changing trend. In this paper, we propose Long Short-Term Preference Modeling for Continuous-Time Sequential Recommendation (\modelname) to capture the evolution of short-term preference under dynamic graph. Specifically, we explicitly encode short-term preference and optimize it via memory mechanism, which has three key operations: \textit{Message}, \textit{Aggregate} and \textit{Update}. Our memory mechanism can not only store one-hop information, but also trigger with new interactions online. Extensive experiments conducted on five public datasets show that \modelname consistently outperforms many state-of-the-art recommendation methods across various lines.\footnote{Our source code will be publicly released.}
\end{abstract}

\section{Introduction}
\label{sec:intro}

Recommender systems have been widely deployed in many online services, such as e-commerce, online review and videos \cite{zhou2018deep,an2019neural}. In sequential recommendation scenario with implicit feedback, modeling the evolution of user preference becomes one of the most essential tasks  \cite{kang2018selfattentive,chang2021sequential}.

Recently, some dynamic graph-based methods \cite{kumar2019predicting,fan2021continuoustime,zhang2022dynamic} for user preference modeling have been studied and achieved state-of-the-art performance by taking both graph structures and timestamps into consideration. For example, TGSRec \cite{fan2021continuoustime} unify sequential patterns and dynamic collaborative signals to improve the quality of recommendation. Despite their superior performance, most of the current methods focus more on user's long-term preference, which can be regarded as user's general/overall preference and remain stable for a long time. However, in real-world recommender systems, user preference keeps evolving, which leads to preference drifting phenomenon \cite{zhou2019deep,chang2021sequential,zheng2022disentangling} and can be regarded as user's short-term preference. 

There exists some works for modeling fast-changing short-term preference in traditional sequential recommendation. They could be roughly summarized into two categories: one attempts to capture short-term preference from user's last few interacted items \cite{li2017neural,liu2018stamp,wu2019sessionbased,kang2018selfattentive}; the other applies time series methods, such as LSTM \cite{graves2012long} and GRU \cite{chung2014empirical}, to capture the evolution of short-term preference from long sequential user behavior data \cite{zhou2019deep,yu2019adaptive,zheng2022disentangling}. 

Although these sequential methods could achieve satisfactory results, how to model short-term preference with dynamic graph-based methods has not been well-addressed yet: (1) The existing dynamic graph-based methods attempt to model user's preference with continuous-time embedding and dynamic collaborative signals, which assume that user's preference evolves smoothly as dynamic graph evolves. Thus, these methods do not explicitly encode and capture the evolution of user's short-term preference as sequential methods do. (2) Each short-term preference has its own evolution track and one behavior of a user may depend on the behavior that takes long time ago \cite{zhou2019deep}. Simply using last few interactions is not enough for modeling the changing trend of short-term preference. Furthermore, both user's long-term and short-term preferences are of great importance for recommendation \cite{liu2018stamp}, but existing dynamic graph-based methods do not combine these two types of preferences well.

Based on all these observations, we propose Long Short-Term Preference Modeling for Continuous-Time Sequential Recommendation (\modelname) to enhance user's short-term preference in user-item dynamic graph. Our method consists of four components: (1) Dynamic Graph Construction that constructs CTDG from history user interaction. (2) Long Short-Term Embedding Layer that explicitly encodes timestamp, long-term and short-term preference with memory mechanism. (3) Dynamic Collaborative Transformer that constructs dynamic embeddings and use DSACF to capture dynamic collaborative signals. (4) Recommendation and Optimization that calculate affinity scores and BPR loss. Specifically, in order to capture the evolution of user's dynamic short-term preference, we explicitly encode short-term preference and propose memory mechanism to optimize it. Inspired by existing memory network \cite{chen2018sequential,pi2019practice} and message passing mechanism, our memory mechanism has three key operations: \textit{Message}, \textit{Aggregate} and \textit{Update}. Firstly, we construct messages according to user/item short-term embedding and continuous-time embedding. Note that such messages take both collaborative signals and time interval information into consideration. Secondly, we aggregate messages from neighbors to obtain information of current preference. Thirdly, we adopt time series method, such as LSTM and GRU, to update short-term embedding with current preference information, thus capturing the evolution of dynamic short-term preference. Our memory mechanism has two distinct advantages: (1) it stores one-hop information during message passing process; (2) it can be triggered with new interactions even after training. In sum, we make the following contributions:

\begin{itemize}[leftmargin=*]
    \item We highlight the evolution of dynamic short-term preference, and take the pioneer of capturing it in dynamic graph-based methods.
    \item We propose \modelname to capture the evolution of short-term preference under dynamic graph. The short-term embeddings that represents such evolution are explicitly encoded and optimized by our proposed memory mechanism. 
    \item We conduct extensive experiments on five public recommendation datasets. Experiments show that our \modelname achieves significant gains over many state-of-the-art baselines from various lines. Further analysis demonstrates the great importance of memory mechanism proposed by our method.
\end{itemize}

\section{Preliminaries}
\label{sec:define}


In this section, we define concepts used in our method. 


\subsubsection{Continuous-Time Dynamic Graph.}
A continuous-time dynamic graph (CTDG) is defined as $\mathcal{G} = \{\mathcal{U}, \mathcal{I}, \mathcal{E}, \mathcal{X}\}$, where $\mathcal{U} = \{u_1, u_2, ..., u_n\}$ is the user node set, $\mathcal{I} = \{i_1, i_2,  ..., i_r\}$ is the item node set, $(u_j, i_o, t_q) \in \mathcal{E}$ represents an interaction between user $u_j$ and item $i_o$ at timestamp $t_q$. $\mathcal{X} = \{ \bm{x}_{j, o}(t_q) | (u_j, i_o, t_q) \in \mathcal{E}\}$ denotes the features of the interactions, which can be optional. $\mathcal{T} = \{t_1, ..., t_p, t_q, ...\}$ is the timestamp set. 



\subsubsection{Dynamic Neighbor Sampling.}\label{neighbor}
Given a user $u_j$ at $t_q$, its neighbors generated by dynamic neighbor sampling are defined as $\mathcal{N}_{u_j}(t_q) = \left\{i_k | (u_j, i_k, t_p) \in \mathcal{E}, t_p < t_q \right\}$. To ensure timeliness, we always sample the most recent $\epsilon$ items for $u_j$ before $t_q$, i.e., $|\mathcal{N}_{u_j}(t_q)| = \epsilon$. 




\subsubsection{Continuous-Time Sequential Recommendation.} 
In this paper, we follow the definition of continuous-time sequential recommendation by \cite{fan2021continuoustime}, in which the input and output can be written as: 

\noindent \textbf{Input:} user set $\mathcal{U}$, item set $\mathcal{I}$, CTDG, timestamp set $\mathcal{T}_u$.

\noindent \textbf{Output:} A continuous-time sequential recommendation that estimates the probability that a user with CTDG  will generate a ranking list of items at $t_c \in \mathcal{T}_u$.

\section{Proposed Method}
\label{sec:proposed}

\begin{figure*}[h]
  \centering
  \includegraphics[width=0.85\linewidth]{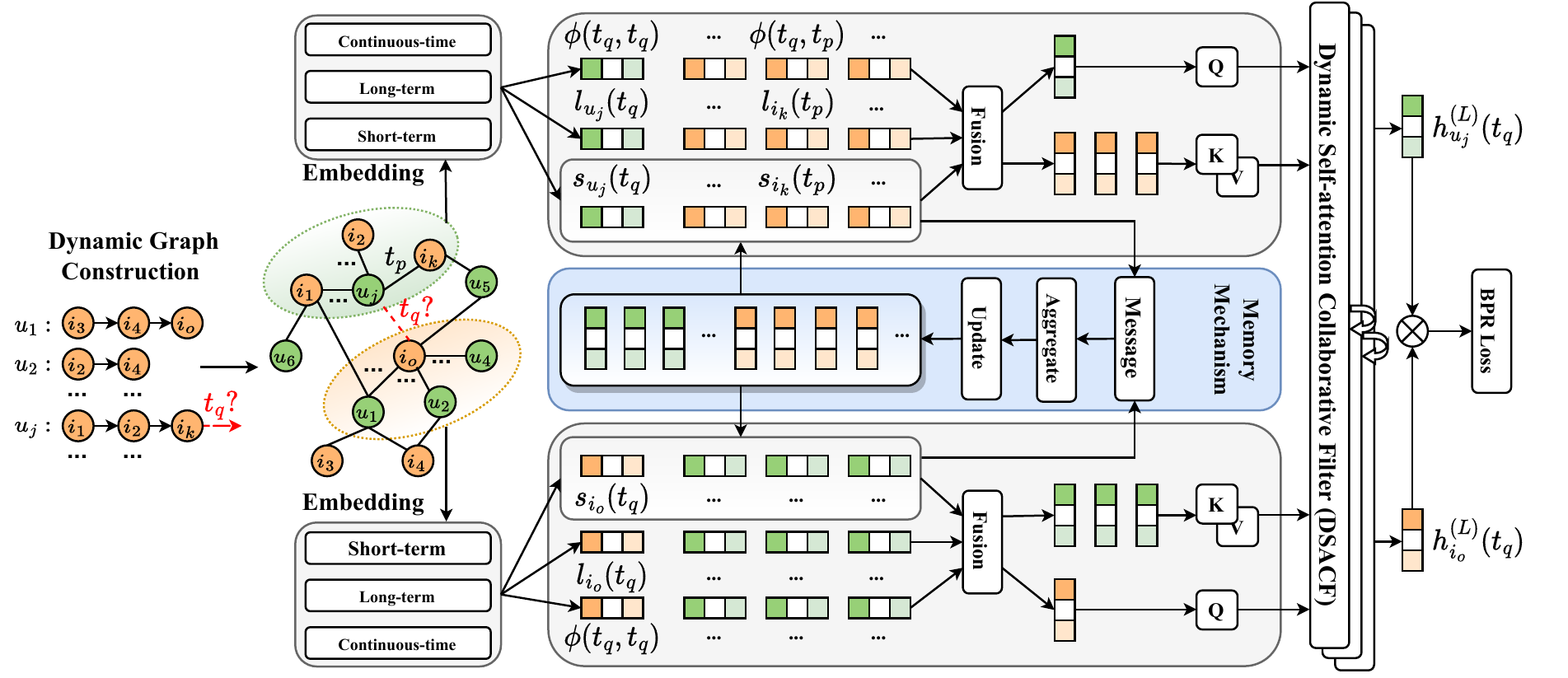}
  \vspace{-0.1in}
  \caption{The method of \modelname. The CTDG is constructed from history user behavior, and the dynamic neighbor sampling is used to generate $\epsilon$ neighbors for $u_j$ and $i_o$, respectively. To obtain final dynamic embedding $\bm{h}^{(L)}_{u_j}(t_q)$ and $\bm{h}^{(L)}_{i_o}(t_q)$, we first generate continuous-time embeddings, long-term embeddings, and enhance short-term embeddings with memory mechanism. Then, we fuse these embeddings to generate query, key, value for DSACF. Finally, we calculate the score between $u_j$ and $i_o$ at $t_q$ and use BPR loss for topk recommendation.}
  \label{tgnrec}
  \vspace{-0.15in}
\end{figure*}

We now present the proposed \modelname method, which is illustrated in Figure \ref{tgnrec}. 
There are four components in the architecture: 
(1) Dynamic Graph Construction: constructs CTDG from history user interaction. 
(2) Long Short-Term Embedding Layer: encodes timestamp as vector, long-term preference as global embedding, and updates short-term embedding by memory mechanism. The memory mechanism is continuous time-aware so that the impact of recent interactions can be adaptively determined. 
(3) Dynamic Collaborative Transformer: constructs dynamic embeddings and use DSACF to capture dynamic collaborative signals.
(4) Recommendation and Optimization: takes user and item representation as input, and outputs a ranked list of items for a given user at a given timestamp.

\subsection{Dynamic Graph Construction} 
We attempt to construct a CTDG from history user behavior, which can be represented by an chronologically ordered list for user $u_j$: $\bm{\zeta}_{u_j} = \{(i_1, t_1), ..., (i_k, t_p), (i_o, t_q), ... \}$. When the user $u_j$ acts on item $i_o$ at $t_q$ in $\bm{\zeta}_{u_j}$, an edge $(u_j, i_o, t_q) \in \mathcal{E}$ is established. The edge feature $\bm{x}_{j, o}(t_q) \in \mathbb{R}^{d_e}$ normally contains information about current interaction, which is determined by dataset. Note that one user can interact with a specific item at any timestamp even many times, which means CTDG is more dense than static graphs.

\subsection{Long Short-Term Embedding Layer} 
We encode three types of embeddings in this paper. The continuous-time embedding module embeds timestamps into vectors. The long-term embedding serves as a global representation of user/item, while the short-term embedding captures the dynamic evolution of user/item with memory mechanism.

\subsubsection{Continuous-Time Embedding.}
We define the dynamic kernel function \cite{xu2020inductive,fan2021continuoustime} to generate continuous-time embeddings of $t_q$ and $t_p$ as: 
\begin{equation}
\label{eq:time 1}
\begin{aligned}
    \phi(t_q, t_p) = \Phi(t_q) \cdot \Phi(t_p)  \in \mathbb{R}^{d_t}
\end{aligned}
\end{equation}
where $\Phi(t): t \rightarrow \mathbb{R}^{d_t}$ is a function mapping based on Bochner's Theorem \cite{loomis2013introduction}. In our method, the function can be formulated as: 
\begin{equation}
\label{eq:time 2}
    \Phi(t) := \sqrt{\frac{1}{d}} [\cos(w_1 t), \sin(w_1 t), ..., \cos(w_{d_t} t), \sin(w_{d_t} t)]
\end{equation}
where $\bm{w} = (w_1, ..., w_{d_t}) \in \mathbb{R}^{d_t}$ denote learnable parameters, which are initialized to uniform distribution. Note that other parameterized functions can also be applied, such as timestamp projection: $\phi_{\Delta}(t_q, t_p) = \mathbf{w}(t_q - t_p)$, where $\mathbf{w}$ are learnable parameters. 

\subsubsection{Long Short-Term User/Item Embedding.}\label{long_short_emb}
Firstly, for long-term embedding, a user (item) node is parameterized by a vector $\bm{l}_{u_j}(t_q) \left( \bm{l}_{i_o}(t_q) \right) \in \mathbb{R}^d$, where we initialize node by indexing an embedding table for user $u_j$ (item $i_o$) with ID embedding: 
\begin{equation}
    \bm{l}_{u_j}(t_q) = \bm{W}_l \cdot \bm{a}_j,\quad \left(\bm{l}_{i_o}(t_q) = \bm{W}_l \cdot \bm{a}_o \right)
\end{equation}
where $\bm{W}_l \in \mathbb{R}^{d\times (n+r)}$ is a linear transformation matrix shared by all users and items. $\bm{a}_j (\bm{a}_o) \in \mathbb{R}^{(n+r)}$ are one-hot ID embeddings. 
As for long-term embedding for users and items, it functions as node features and is optimized by gradient decent.

Then, for short-term embedding, a user (item) is parameterized by a vector $\bm{s}_{u_j}(t_q) \left( \bm{s}_{i_o}(t_q) \right) \in \mathbb{R}^d$, which is initialized to all-zero vectors, and optimized by memory mechanism.



\subsubsection{Memory Mechanism.}\label{short update}
We propose memory mechanism to model the dynamic evolution of short-term preference over time. Owning to memory mechanism, \modelname has the capability to memorize the compressed node history. The memory mechanism, which is updated if a node has new interactions (even after training), has three major operations: \textit{Message}, \textit{Aggregate} and \textit{Update}. In the following, we take user $u_j$ at $t_q$ as an example, while item $i_o$ has similar operations. 

\noindent \textbf{Message}. Since dynamic collaborative signals and time interval information are essential in recommendation scenarios, we combine user/item short-term embeddings and continuous-time embedding to construct the messages. For each user $u_j$ at timestamp $t_p$, we can compute the message from its neighbor $i_k \in \mathcal{N}_{u_j}(t_q)$ as: 
\begin{equation}
\label{eq:short 1}
    \bm{m}_{u_j \leftarrow i_k}(t_p) = \left[\bm{s}_{u_j}(t_q^-) \| \bm{s}_{i_k}(t_p^-) \| \bm{x}_{j, k}(t_p) \| \phi(t_q, t_p) \right]
\end{equation}
where $\|$ is the concatenate operation. $\bm{s}_{u_j}(t_q^-)$ and $\bm{s}_{i_k}(t_p^-)$ represent the latest states of short-term embeddings for $u_j$ and $i_k$ right before $t_q$ and $t_p$, respectively. $\phi(t_q, t_p)$ represents continuous-time embeddings.

\noindent \textbf{Aggregate}. In this stage, we aggregate the messages from the neighbors of $u_j$ at timestamp $t_q$ to obtain current preference information. The aggregation operation can be formulated as:
\begin{equation}
\label{eq:short 2}
    \mathbf{\overline{m}}_{u_j}(t_q) = \mathrm{Agg}\left\{\bm{m}_{u_j \leftarrow i_k}(t_p), i_k \in \mathcal{N}_{u_j}(t_q) \right\}
\end{equation}
where $\mathrm{Agg}\{\cdot\}$ is an aggregation function. While multiple choices can be considered for implementing aggregation function, for example, mean or attention. In our method, we use last time aggregation, which selects the message from the last interaction before $t_q$. The motivation is that user's preference may evolve dynamically, therefore, the most recent message might be more valuable than other messages. 

\noindent \textbf{Update}. Finally, to capture the evolution of dynamic short-term preference, we update short-term embedding $s_{u_j}(t_q)$ based on $\mathbf{\overline{m}}_{u_j}(t_q)$ and $s_{u_j}(t_q^-)$:
\begin{equation}
\label{eq:short 3}
    \bm{s}_{u_j}(t_q) = f\left(\mathbf{\overline{m}}_{u_j}(t_q), \bm{s}_{u_j}(t_q^-)\right)
\end{equation}
The update function $f(\cdot)$ can be a time series function, such as RNN \cite{medsker2001recurrent}, LSTM \cite{graves2012long}, GRU \cite{chung2014empirical}, etc. In this paper, we implement $f(\cdot)$ as GRU. Analogously, we can obtain the embedding $\bm{s}_{i_k}(t_q)$ for item $i_k$ by the above three operations. 

To summarize, the reason that memory mechanism boosts the performance lies in its message passing process of updating short-term embeddings for users and items. Since message passing process in memory mechanism has already stored 1-hop messages, \modelname only needs one layer to achieve good results. (Our method actually aggregates the information of 2-order neighbors.)

\subsection{Dynamic Collaborative Transformer (DCT)}
Next, we present dynamic collaborative transformer (DCT) layer, which contains two component: (1) dynamic embedding construction (2) dynamic self-attention collaborative filter (DSACF). In the following, we take the calculation of final dynamic embedding for $u_j$ at $t_q$ as an example.

\subsubsection{Dynamic Embedding Construction.}\label{time_encode}

We construct the dynamic embedding $\bm{z}_{u_j}^{(\ell)}(t_q)$ for user $u_j$ at $t_q$ of the $\ell$-th DCT layer as the combination of long short-term information and continuous-time embeddings: 
\begin{equation}\label{hu}
    \bm{z}_{u_j}^{(\ell)}(t_q) = \bm{h}_{u_j}^{(\ell)}(t_q)  \| \phi(t_q, t_q)
\end{equation}
where $\ell = \{0, ..., L-1\}$ denotes the layer of DCT. $\phi(t_q, t_q)$ denotes continuous-time embedding. $\bm{h}_{u_j}^{(\ell)}(t_q)$ is the long short-term information for $u_j$ at $t_q$. Note that when $\ell=0$, it is the first DCT layer. The long short-term information $\bm{h}_{u_j}^{(0)}(t_q) = \bm{s}_{u_j}(t_q) + \bm{l}_{u_j}(t_q)$, i.e., the summation of long-term and short-term embeddings. When $\ell > 0$, the long short-term information is generated from the previous DCT layer.

Different from user $u_j$, the dynamic embedding for its neighbor $i_k \in \mathcal{N}_{u_j}(t_q)$ at $t_p$ contains edge features $\bm{x}_{j, k}(t_p)$, which can be formulated as:
\begin{equation}\label{eq7}
    \bm{z}_{i_k}^{(\ell)}(t_p) = \bm{h}_{i_k}^{(\ell)}(t_p) \| \bm{x}_{j, k}(t_p) \| \phi(t_q, t_p)
\end{equation}
Again, note that when $\ell=0$, $\bm{h}_{i_k}^{(0)}(t_p) = \bm{s}_{i_k}(t_p) + \bm{l}_{i_k}(t_p)$. When $\ell > 0$, the long short-term information is output from the previous DCT layer.

\subsubsection{Dynamic Self-attention Collaborative Filter (DSACF).}
Next, we build upon the masked self-attention architecture \cite{vaswani2017attention} in order to capture dynamic collaborative signal. We compute the weighted summation of the information from all sampled items as:
\begin{equation}\label{znu}
    \bm{z}_{\mathcal{N}_{u_j}}^{(\ell)}(t_q) =  \sum_{i_k \in \mathcal{N}_{u_j}(t_q)}\mathcal{\alpha}_{j, k}(t_p) \cdot \bm{W}_v^{(\ell)} \cdot \bm{z}_{i_k}^{(\ell)}(t_p)
\end{equation}
where $\alpha_{j, k}(t_p)$ denotes the weights of the interaction and $\bm{W}_v^{(\ell)} \in \mathbb{R}^{d\times (d+d_t+d_e)}$ is the linear transformation matrix. 
The attention weights $\alpha_{j, k}(t_p)$ can be formulated as follows:

\begin{equation}\label{alpha}
    \alpha_{j, k}(t_p) = \frac{\exp \left(\beta_{j, k}(t_p) \right)}{\sum_{i_{k'} \in \mathcal{N}_{u_j}(t_q)}\exp(\beta_{j, k'}(t_p))}
\end{equation}
where $\beta_{j, k}(t_p)$ represents the raw attention weights, which can be calculated as: %
\begin{equation}\label{beta}
    \beta_{j, k}(t_p) = \frac{1}{\sqrt{d+d_t}} \left(\bm{W}_k^{(\ell)} \cdot \bm{z}_{i_k}^{(\ell)}(t_p) \right)^T \left(\bm{W}_q^{(\ell)} \cdot \bm{z}_{u_j}^{(\ell)}(t_q) \right)
\end{equation}
where $\bm{W}_{k}^{(\ell)} \in \mathbb{R}^{d\times (d+d_t+d_e)}$ and $\bm{W}_{q}^{(\ell)} \in \mathbb{R}^{d\times (d+d_t)}$ are both linear transformation matrices. 
Finally, we combine $\bm{z}_{\mathcal{N}_{u_j}}^{(\ell)}(t_q)$ and the input dynamic embedding at current DCT layer, then forward it to a FFN \cite{sazli2006brief} layer: 
\begin{equation}
    \bm{h}_{u_j}^{(\ell+1)}(t_q) = \mathrm{FFN} \left(\bm{z}_{\mathcal{N}_{u_j}}^{(\ell)}(t_q)\|\bm{z}_{u_j}^{(\ell)}(t_q)\right)
\end{equation}
As for $(u_j, i_o, t_q)$, in the last layer $L - 1$, we can obtain the final dynamic embedding $\bm{h}_{u_j}^{(L)}(t_q)$ for $u_j$ at $t_q$. Analogously, $\bm{h}_{i_o}^{(L)}(t_q)$ for item $i_o$ can also be calculated by DCT. In addition, we employ multi-head mechanism to our \modelname, which not only stabilize the learning process of self-attention, but also be beneficial to performance for continuous-time sequential recommendation.


\subsection{Recommendation and Optimization}
\textbf{Prediction Layer}.
For each $(u_j, i_o, t_q)$, we can calculate the prediction score as: 
\begin{equation}
\begin{aligned}
    \bm{y}_{j, o}(t_q) = \mathrm{FFN} \left(\bm{h}_{u_j}^{(L)}(t_q) \| \bm{h}_{i_o}^{(L)}(t_q) \right) \\
\end{aligned}
\end{equation}
where $\bm{y}_{j, o}(t_q)$ denotes the affinity score between user $u_j$ and item $i_o$ at $t_q$. Note that other operations can also be used, such as hadamard product \cite{horn1990hadamard}. The affinity score is used to calculate the ranking list of items for each user, while items with high scores will be recommended for user at the top.

\noindent \textbf{Loss Function}.
We use the pairwise BPR loss \cite{rendle2012bpr} to learn the parameters. BPR loss has been widely used in recommender systems and assumes that the positive interactions should be assigned higher affinity scores than the negative ones. The objective function is as follows:  
\begin{equation}
    \mathcal{L} = \sum_{u_j} \sum_{(u_j, i_o, i_{o'}, t_q) \in \mathcal{O}} -\ln \sigma \left(\bm{y}_{j, o}(t_q) - \bm{y}_{j, {o'}}(t_q)\right) + \lambda \| \bm{\Theta} \|_2^2
\end{equation}
where $\mathcal{O} = \{(u_j, i_o, i_{o'}, t_q) | (u_j, i_o, t_q) \in \mathcal{E} , (u_j, i_{o'}, t_q) \notin \mathcal{E}\}$ denotes the pairwise training data; $\sigma(\cdot)$ is the sigmoid function; $\bm{\Theta}$ represents the weights of long-term embeddings, and $\lambda$ controls the $L_2$ regularization strength to prevent overfitting. 

\section{Experiment}
\label{sec:experiment}

In this section, we perform experiments on five public datasets to demonstrate the effectiveness of \modelname. The experiment analysis answers the following Research Questions (RQs):

\begin{itemize}[leftmargin=*]
    \item \textbf{RQ1:} How does \modelname perform as compared with state-of-the-art methods in continuous-time sequential recommendation task?
    \item \textbf{RQ2:} Dose \modelname capture users' long-term and short-term preference effectively?
    \item \textbf{RQ3:} How do different modules affect the performance of \modelname?
    \item \textbf{RQ4:} Does \modelname have efficient training and inference process?
\end{itemize}


\subsection{Datasets}\label{dataset}

We conduct our experiments on five datasets, which are publicly accessible and cover various domain and sparsity. The datasets\footnote{\url{http://snap.stanford.edu/jodie}} of Wikipedia and Reddit are released by JODIE \cite{kumar2019predicting}: (1) Wikipedia dataset contains one month of user-edit interactions on Wikipedia pages. (2) Reddit dataset contains one month of user-post interactions in Sub-Reddit. We also use Amazon review dataset\footnote{\url{http://jmcauley.ucsd.edu/data/amazon}} \cite{mcauley2015image}, which contains product reviews and metadata from Amazon. We choose two subsets, i.e., Software, Musical Instruments, and Luxury Beauty for experiments. The statistical information is shown in Table \ref{tab:datasets}.

For each dataset, in main experiment, we chronologically split it into train/validation/test at 80\%/10\%/10\% ratio. Moreover, to verify the effectiveness of long short-term preference modeling, we set up more data partitioning schemes. We create interaction sequences of varying length for training. We take the latest 25\%, 50\%, 75\%, and 100\% interactions of the full training set, respectively. The validation set and test set are the same as in main experiment.

\begin{table}[h]
\vspace{-0.15in}
  \caption{Statistics of datasets. Density indicates the graph density \cite{anderson1999interaction}. 
  }
  \label{tab:datasets}
  \resizebox{\linewidth}{!}{
  \begin{tabular}{l|c|c|c|c|c}
    \toprule
    Dataset & Wikipedia & Reddit & Software & Instruments & Beauty \\ 
    \midrule
    \#Users & 8,227 & 10,000 & 1,827 & 1,430 & 3,819 \\
    \#Items & 1,000 & 984 & 803 & 900 & 1,581 \\
    \#Instances & 157,474 & 672,447 & 12,805 & 10,261 & 34,278 \\
    \#Edge features & 172 & 172 & -- & -- & -- \\
    Density & 0.36\% & 1.11\% & 0.37\% & 0.37\% & 0.23\%\\
  \bottomrule
\end{tabular}
}
\vspace{-0.15in}
\end{table}

\subsection{Baselines}
To demonstrate the effectiveness, we compare our \modelname with the existing three categories of seven state-of-the-art methods:

\begin{itemize}[leftmargin=*]
    \item \textbf{General recommendation methods:} In this part, we compare with \textbf{NGCF} \cite{wang2019neural} and \textbf{LightGCN} \cite{he2020lightgcn}, which are the state-of-the-art graph-based methods in general recommendation. For fair comparison, we extend NGCF to an edge feature-aware version NGCF$\dagger$\footnote{More details will be released in our code.\label{edge_version}}, in which the edge features are added into its message propagation.
    Note that LightGCN is not extended as the above, as it does not have a weight matrix. 
    
    \item \textbf{Sequential recommendation methods:} In this part, we compare with \textbf{SASRec} \cite{kang2018selfattentive}, \textbf{TiSASRec} \cite{li2020time} and \textbf{TGSRec} \cite{fan2021continuoustime}, which are the state-of-the-art sequential recommendation methods. Similar to NGCF, we also extend them into the edge feature-aware version: SASRec$\dagger$, TiSASRec$\dagger$ and TGSRec$\dagger$\textsuperscript{\ref{edge_version}}. 
    
    \item \textbf{Dynamic graph embedding methods:} In this part, we compare with \textbf{JODIE} \cite{kumar2019predicting} and \textbf{DyREP} \cite{trivedi2019dyrep}, which are the state-of-the-art dynamic graph embedding methods. 
\end{itemize}

\subsection{Experimental Details}
\subsubsection{Evaluation Setting.}
To evaluate the quality of recommended items, we adopt two metrics \cite{he2017neural,hsieh2017collaborative}: Recall@K, NDCG@K. We set K in \{10, 20\} for a comprehensive comparison. 
We report the average and standard deviation of resluts over 10 independent runs. For each interaction $(u_j, i_o, t_q)$ in validation/test sets, we treat items that $u_j$ has not interacted with before $t_q$ as negative items. 
In order to accelerate the evaluation, we sample 500 negative items for evaluation instead of full set of negative items in Wikipedia, Reddit and Instruments.


\subsubsection{Parameter Settings.}
We implement our \modelname method with PyTorch \cite{paszke2019pytorch} in a NVIDIA Tesla V100 GPU (32GB). In Amazon, we fixed the time embedding size, long and short-term embedding sizes to 172, while in Instruments, it is fixed to 128. We optimize all methods with the Adam optimizer, where the batch size is fixed at 200. We tune the learning rate in $\{10^{-3}$, $10^{-4}$, $10^{-5}\}$, search the $L_2$ regularization coefficient from $\{10^{-1}$, $3 \times 10^{-1}$, $10^{-2}$, $10^{-3}\}$. We also search for number of layers in \{1, 2, 3\}, and number of heads in \{2, 4\}. We use the xavier\_uniform initializer \cite{glorot2010understanding} to initialize the method parameters.

\begin{table*}[h]
  \caption{The mean and standard deviation over 10 different runs on five datasets. \ul{Underline} means the best baseline, \textbf{bold} means the best performance. The Improv. indicates the improvement of \modelname compared with the best baseline. $\dagger$ means the method uses its edge feature-aware version in Wikipedia and Reddit dataset.}
  \label{tab:all}
  \resizebox{\linewidth}{!}{
  \begin{tabular}{c|c|cccccccc|r}
    \toprule
    Datasets & Metrics & \multicolumn{1}{c}{NGCF$\dagger$} & \multicolumn{1}{c}{LightGCN} & \multicolumn{1}{c}{SASRec$\dagger$} & \multicolumn{1}{c}{TiSASRec$\dagger$} & \multicolumn{1}{c}{JODIE} & \multicolumn{1}{c}{DyREP} & \multicolumn{1}{c}{TGSRec$\dagger$} & \multicolumn{1}{c}{\textbf{\modelname}} & \multicolumn{1}{|c}{Improv.}  \\
    \midrule
    \multirow{5}{*}{Wikipedia} & Recall@10 & 0.0292\std{0.0036} & 0.0262\std{0.0041} & 0.5319\std{0.0184} & 0.5350\std{0.0156} & 0.4589\std{0.0219} & 0.5707\std{0.0156} & \ul{0.7488}\std{0.0052} & \textbf{0.8602}\std{0.0028} & +0.1114 \\
    ~ & Recall@20 & 0.0438\std{0.0069} & 0.0394\std{0.0061} & 0.5920\std{0.0157} & 0.5991\std{0.0119} & 0.5532\std{0.0206} & 0.6294\std{0.0139} & \ul{0.8012}\std{0.0051} & \textbf{0.8859}\std{0.0027} & +0.0847 \\
    ~ & NDCG@10 & 0.0340\std{0.0046} & 0.0304\std{0.0035} & 0.3694\std{0.0097} & 0.3717\std{0.0066} & 0.3333\std{0.0206} & 0.4791\std{0.0131} & \ul{0.6420}\std{0.0055} & \textbf{0.8045}\std{0.0034} & +0.1625 \\
    ~ & NDCG@20 & 0.0413\std{0.0055} & 0.0370\std{0.0039} & 0.3847\std{0.0088} & 0.3880\std{0.0052} & 0.3568\std{0.0191} & 0.4937\std{0.0129} & \ul{0.6551}\std{0.0054} & \textbf{0.8109}\std{0.0031}  & +0.1558 \\
    \midrule
    \multirow{4}{*}{Reddit} & Recall@10 & 0.0221\std{0.0006} & 0.0218\std{0.0004} & 0.7699\std{0.0021} & 0.7700\std{0.0009} & 0.3053\std{0.0405} & 0.7656\std{0.0044} & \ul{0.8005}\std{0.0035} & \textbf{0.8306}\std{0.0023} & +0.0301 \\
    ~ & Recall@20 & 0.0331\std{0.0005} & 0.0324\std{0.0006} & 0.8282\std{0.0023} & 0.8287\std{0.0014} & 0.3910\std{0.0483} & 0.8188\std{0.0036} & \ul{0.8439}\std{0.0031} & \textbf{0.8697}\std{0.0021} & +0.0258 \\
    ~ & NDCG@10 & 0.0500\std{0.0010} & 0.0487\std{0.0009} & 0.6123\std{0.0023} & 0.6109\std{0.0019} & 0.2126\std{0.0324} & 0.6524\std{0.0061} & \ul{0.7164}\std{0.0049} & \textbf{0.7552}\std{0.0028} & +0.0388 \\
    ~ & NDCG@20 & 0.0608\std{0.0009} & 0.0592\std{0.0013} & 0.6271\std{0.0022} & 0.6258\std{0.0020} & 0.2339\std{0.0340} & 0.6657\std{0.0058} & \ul{0.7273}\std{0.0048} & \textbf{0.7650}\std{0.0027} & +0.0377 \\
    \midrule
    \multirow{4}{*}{Software} & Recall@10 & 0.0882\std{0.0091} & \ul{0.1204}\std{0.0101} & 0.0446\std{0.0019} & 0.0452\std{0.0052} & 0.0943\std{0.0304} & 0.0729\std{0.0148} & 0.0541\std{0.0093} & \textbf{0.1591}\std{0.0144} & +0.0387 \\
    ~ & Recall@20 & 0.1391\std{0.0133} & \ul{0.1785}\std{0.0091} & 0.0782\std{0.0043} & 0.0763\std{0.0051} & 0.1578\std{0.0320} & 0.1284\std{0.0281} & 0.0775\std{0.0074} & \textbf{0.2334}\std{0.0172} & +0.0549 \\
    ~ & NDCG@10 & 0.0415\std{0.0027} & \ul{0.0554}\std{0.0032} & 0.0209\std{0.0019} & 0.0215\std{0.0026} & 0.0538\std{0.0220} & 0.0425\std{0.0066} & 0.0263\std{0.0041} & \textbf{0.0935}\std{0.0044} & +0.0381 \\
    ~ & NDCG@20 & 0.0523\std{0.0043} & 0.0683\std{0.0034} & 0.0293\std{0.0018} & 0.0293\std{0.0018} & \ul{0.0696}\std{0.0194} & 0.0563\std{0.0093} & 0.0321\std{0.0037} & \textbf{0.1120}\std{0.0039} & +0.0424 \\
    \midrule
    \multirow{4}{*}{Instruments} & Recall@10 & 0.0498\std{0.0070} & 0.0484\std{0.0036} & 0.0738\std{0.0044} & \textbf{0.0781}\std{0.0040} & 0.0463\std{0.0109} & 0.0419\std{0.0029} & 0.0604\std{0.0102} & \ul{0.0776}\std{0.0097} & -- \\
    ~ & Recall@20 & 0.0797\std{0.0099} & 0.0769\std{0.0057} & 0.1212\std{0.0073} & \ul{0.1329}\std{0.0043} & 0.0855\std{0.0104} & 0.0831\std{0.0132} & 0.0962\std{0.0086} & \textbf{0.1384}\std{0.0062} & +0.0055 \\
    ~ & NDCG@10 & 0.0188\std{0.0019} & 0.0197\std{0.0018} & 0.0362\std{0.0019} & \ul{0.0408}\std{0.0022} & 0.0229\std{0.0053} & 0.0219\std{0.0008} & 0.0351\std{0.0044} & \textbf{0.0467}\std{0.0041} & +0.0059 \\
    ~ & NDCG@20 & 0.0243\std{0.0023} & 0.0254\std{0.0021} & 0.0481\std{0.0024} & \ul{0.0545}\std{0.0008} & 0.0327\std{0.0050} & 0.0323\std{0.0030} & 0.0441\std{0.0035} & \textbf{0.0620}\std{0.0027} & +0.0075 \\
    \midrule
    \multirow{4}{*}{Beauty} & Recall@10 & 0.0493\std{0.0034} & 0.0548\std{0.0025} & 0.0440\std{0.0109} & 0.0442\std{0.0181} & \ul{0.0890}\std{0.0054} & 0.0519\std{0.0041} & 0.0546\std{0.0086} & \textbf{0.1343}\std{0.0156} & +0.0453 \\
    ~ & Recall@20 & 0.0725\std{0.0046} & 0.0770\std{0.0019} & 0.0745\std{0.0156} & 0.0734\std{0.0227} & \ul{0.1219}\std{0.0074} & 0.0840\std{0.0042} & 0.0860\std{0.0091} & \textbf{0.1852}\std{0.0292} & +0.0633  \\
    ~ & NDCG@10 & 0.0209\std{0.0011} & 0.0222\std{0.0010} & 0.0218\std{0.0051} & 0.0230\std{0.0100} & \ul{0.0544}\std{0.0018} & 0.0319\std{0.0033} & 0.0315\std{0.0055} & \textbf{0.0817}\std{0.0069} & +0.0273 \\
    ~ & NDCG@20 & 0.0253\std{0.0016} & 0.0265\std{0.0012} & 0.0295\std{0.0063} & 0.0303\std{0.0112} & \ul{0.0627}\std{0.0021} & 0.0399\std{0.0033} & 0.0394\std{0.0053} & \textbf{0.0946}\std{0.0096} & +0.0319 \\
  \bottomrule
\end{tabular}
}
\vspace{-0.15in}
\end{table*}

\subsection{Comparison with Baselines (RQ1)}
We compare the performance of baseline methods with \modelname. The results (Table \ref{tab:all}) indicate that:
\begin{itemize}[leftmargin=*]
   \item Our method \modelname yields the best performance on almost all the datasets. In particular, \modelname improves over the strongest baselines \textit{w.r.t.} NDCG@10 by 25.31\%, 5.26\%, 71.87\%, 14.46\% and 50.18\% in five datasets, respectively. Several factors together contribute to the superiority of \modelname: (1) \modelname explicitly models long short-term preference. (2) Memory mechanism enhances short-term preference modeling. (3) Dynamic Collaborative Filter Modeling captures dynamic collaborative signals.
   
   \item Compared to TGSRec, our \modelname enhances short-term preference by memory mechanism, thus leading to high performance in Wikipedia and Reddit datasets. In addition, LightGCN and TiSASRec performs well in Software and Instruments, respectively. This emphasizes the importance of collaborative signals in recommender systems. We also observe that our \modelname has a higher improvement in NDCG@K and MRR metrics, which indicates that our method performs better in ranking.
\end{itemize}

\subsection{Long Short-term Preferences (RQ2)}

\subsubsection{Effect of different training data proportions.}
In a real-world recommeder system, the length of interaction history available for training models may vary. Longer history provides richer information of dynamic patterns than shorter one, but introduces increased noise too. Thus, whether to effectively model the user’s long short-term preference in a short timespan is a significant issue. We study how \modelname improves the recommendation for those users with different training data proportions.

\begin{figure}[h]
\centering
\subfigure[Recall@10 in Wikipedia.]{
\includegraphics[width=0.45\linewidth]{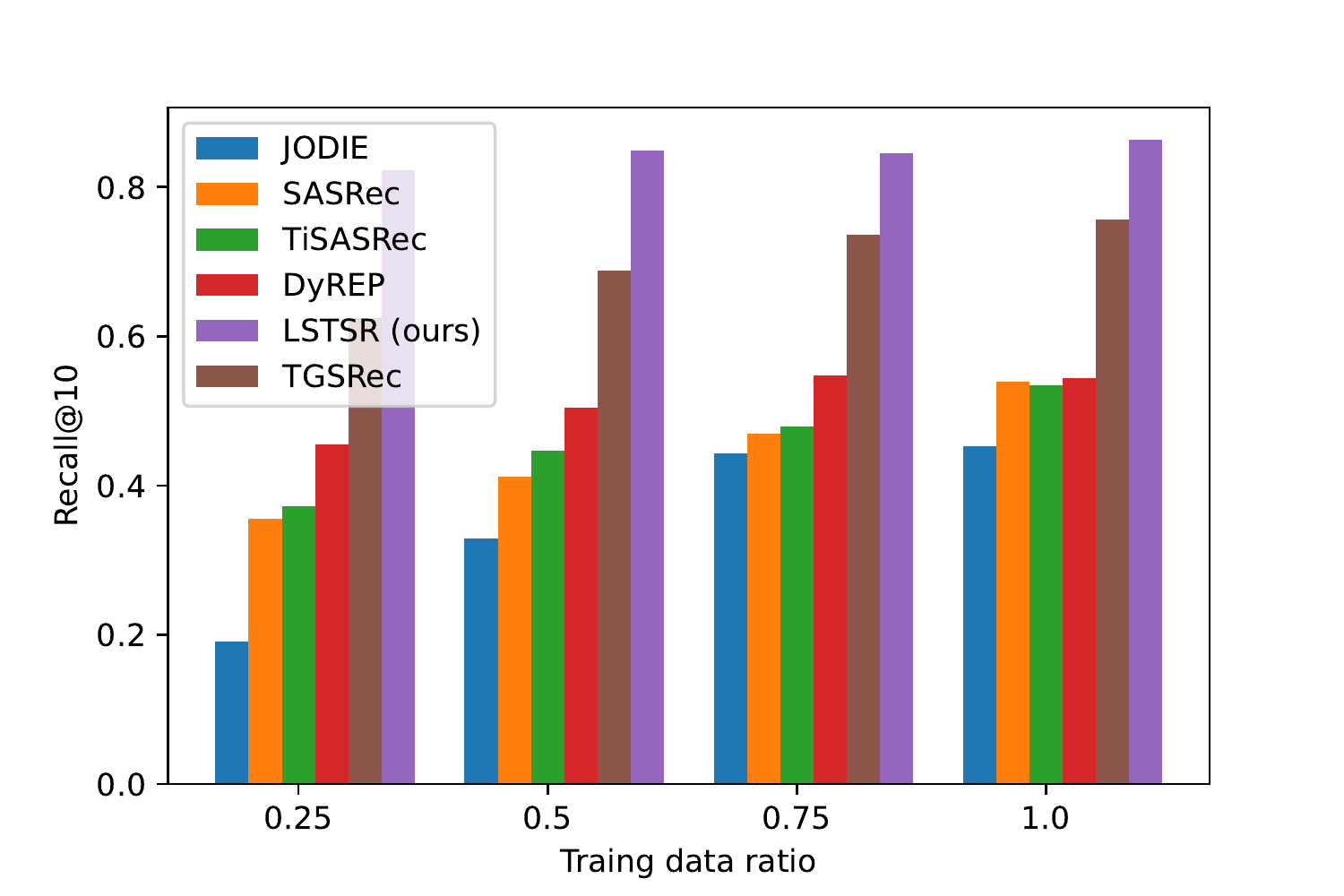}
}
\quad
\subfigure[NDCG@10 in Wikipedia.]{
\includegraphics[width=0.45\linewidth]{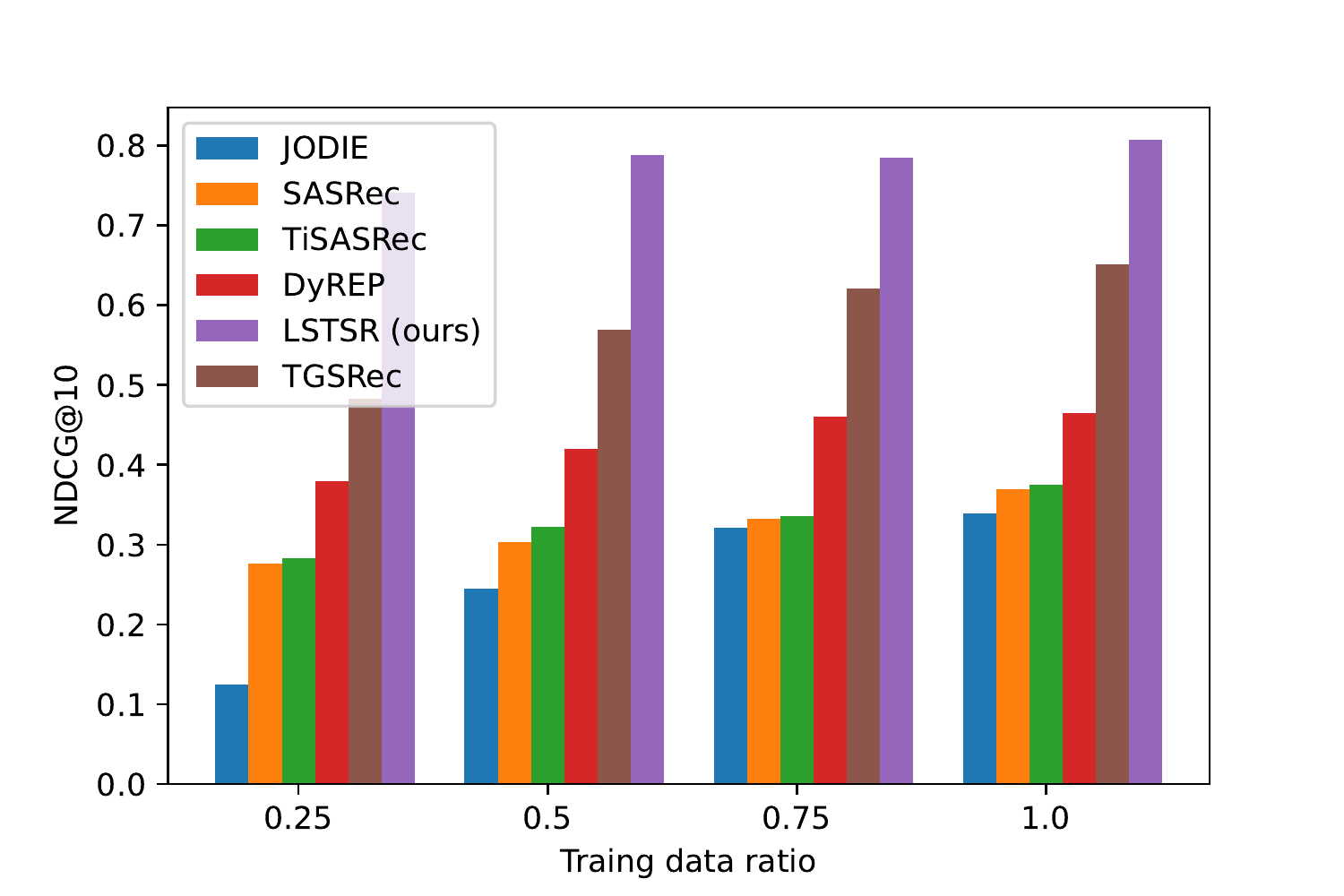}
}
\quad
\subfigure[Recall@10 in Reddit.]{
\includegraphics[width=0.45\linewidth]{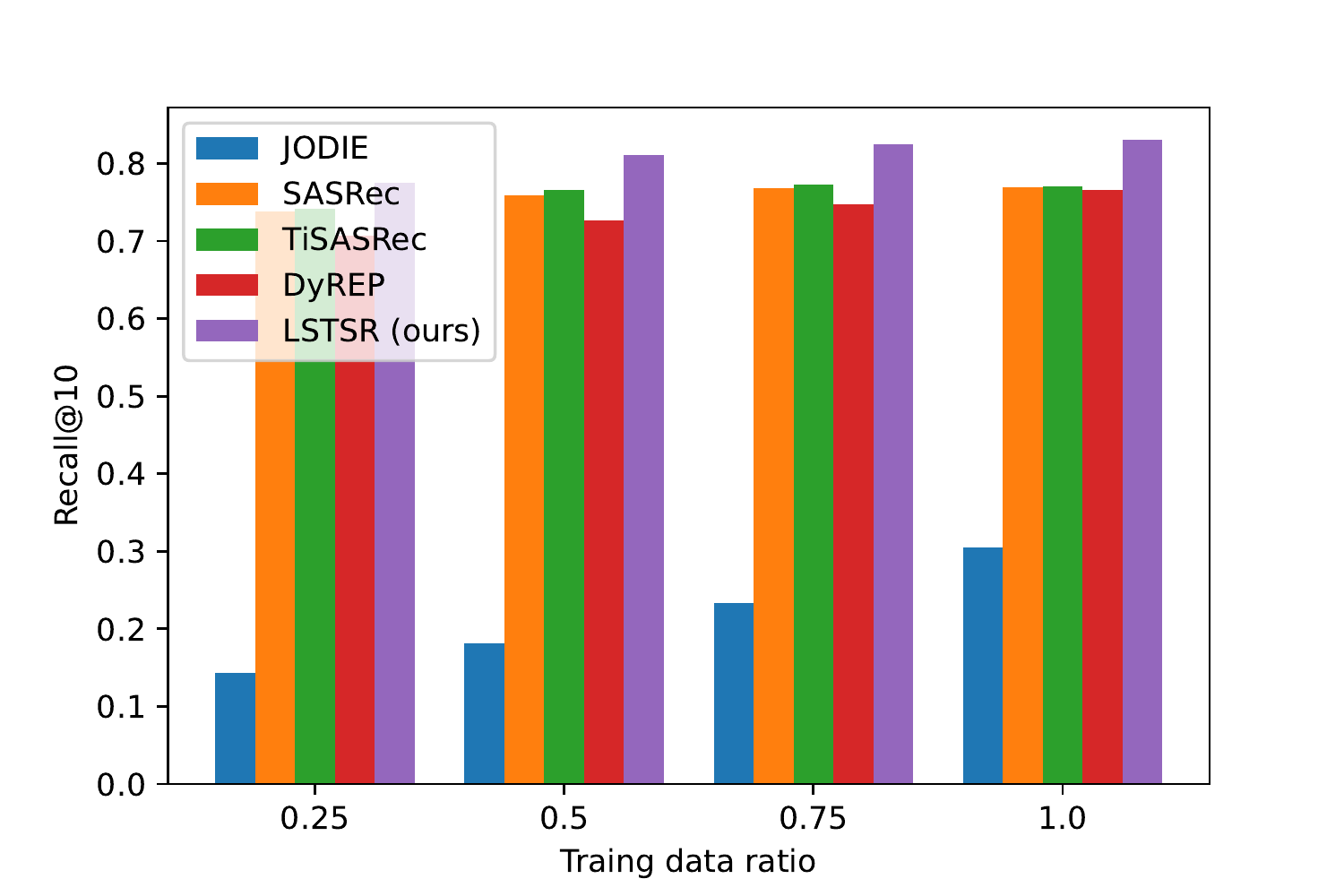}
}
\quad
\subfigure[NDCG@10 in Reddit.]{
\includegraphics[width=0.45\linewidth]{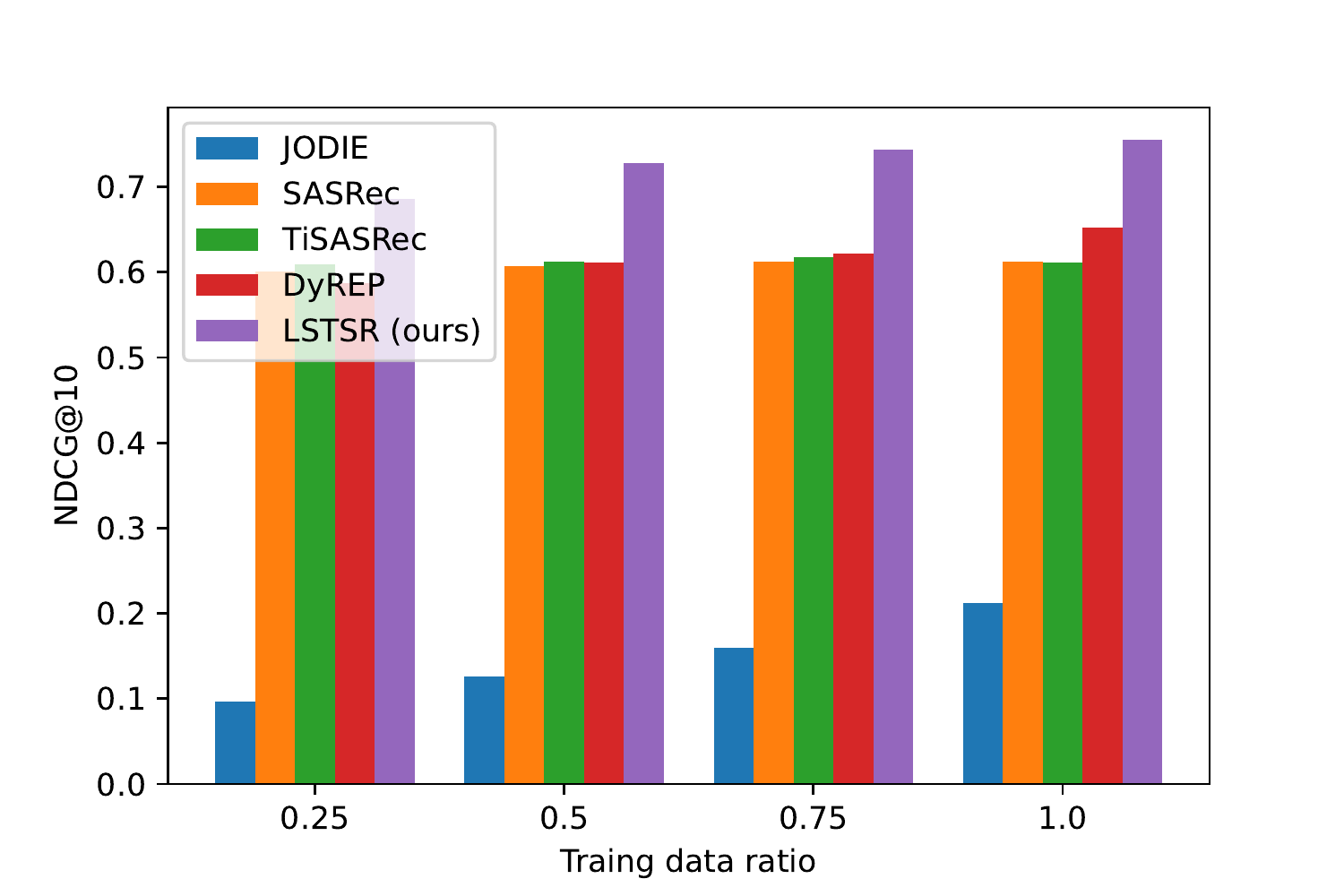}
}
\vspace{-0.15in}
\caption{Long-term and short-term results with different training data proportions on Wikiepdia and Reddit. Result of TGSRec on Reddit dataset is omitted due to long running time ($\ge$ 7 days).}
\label{fig:long short plot}
\vspace{-0.15in}
\end{figure}

As shown in Figure \ref{fig:long short plot}, we can observe that: (1) all baselines have poor performance in capturing users’ long short-term preference when the timespan is short due to data sparsity. However, our method still keeps the excellent performance of NDCG@10 in 0.8076 and 0.7552 on Wikipedia and Reddit with shortest timespan, respectively. 
Since \modelname models the short-term interest explicitly and saves the compressed history of the node information in memory, it can achieve good performance for users with a short timespan. 
(2) As the timespan increases, all methods’ performance improves. Our method \modelname is consistently superior to other methods across two datasets. Meanwhile, TGSRec also get good performance improves. Since our method and TGSRec adopt continuous-time embedding and dynamic collaborative filter methods, they can learn good performance for users.

\begin{figure}[ht]
\centering
\subfigure[Training set for user-320.]{
\includegraphics[width=0.47\linewidth]{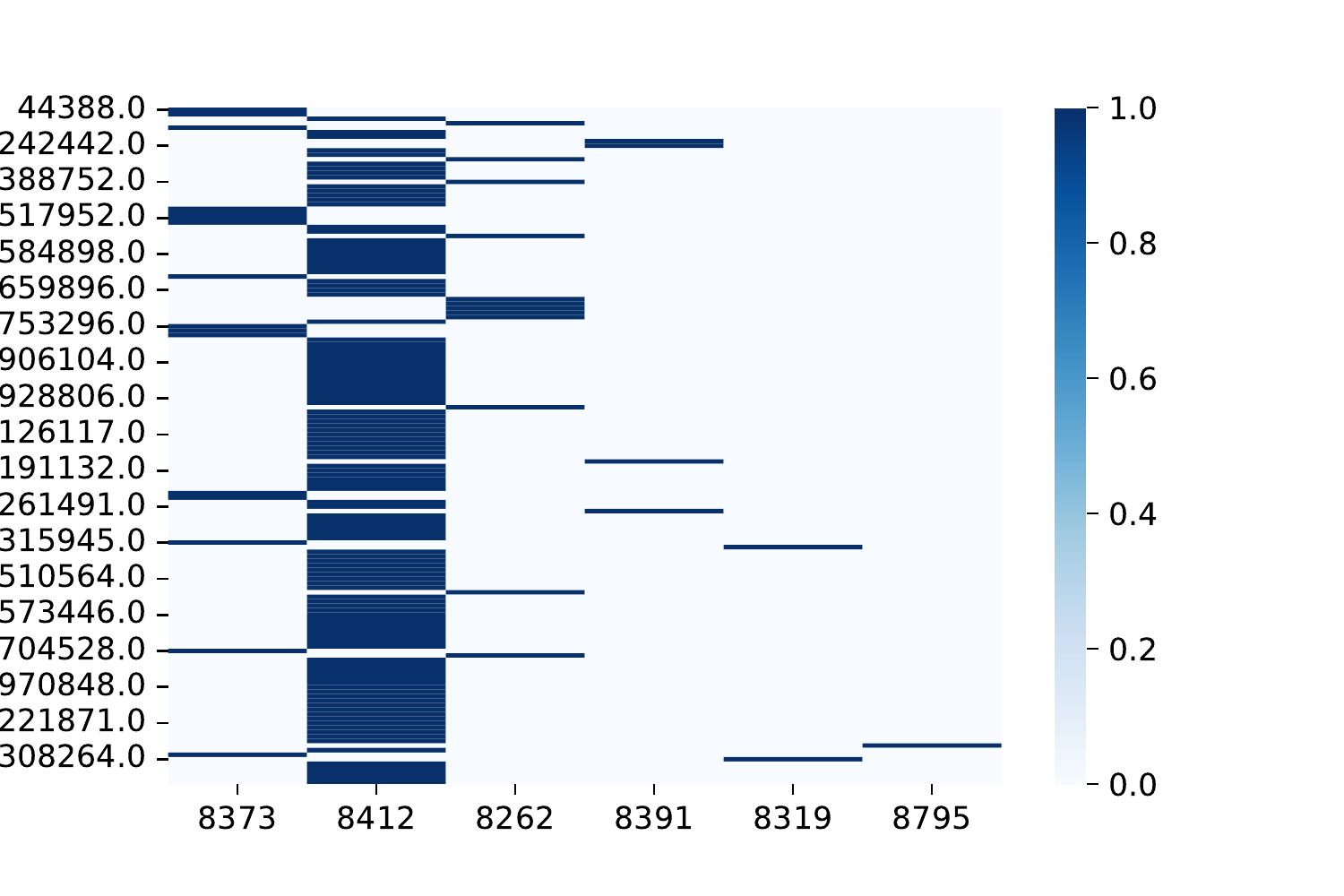}
}
\subfigure[Predictions for user-320.]{
\includegraphics[width=0.47\linewidth]{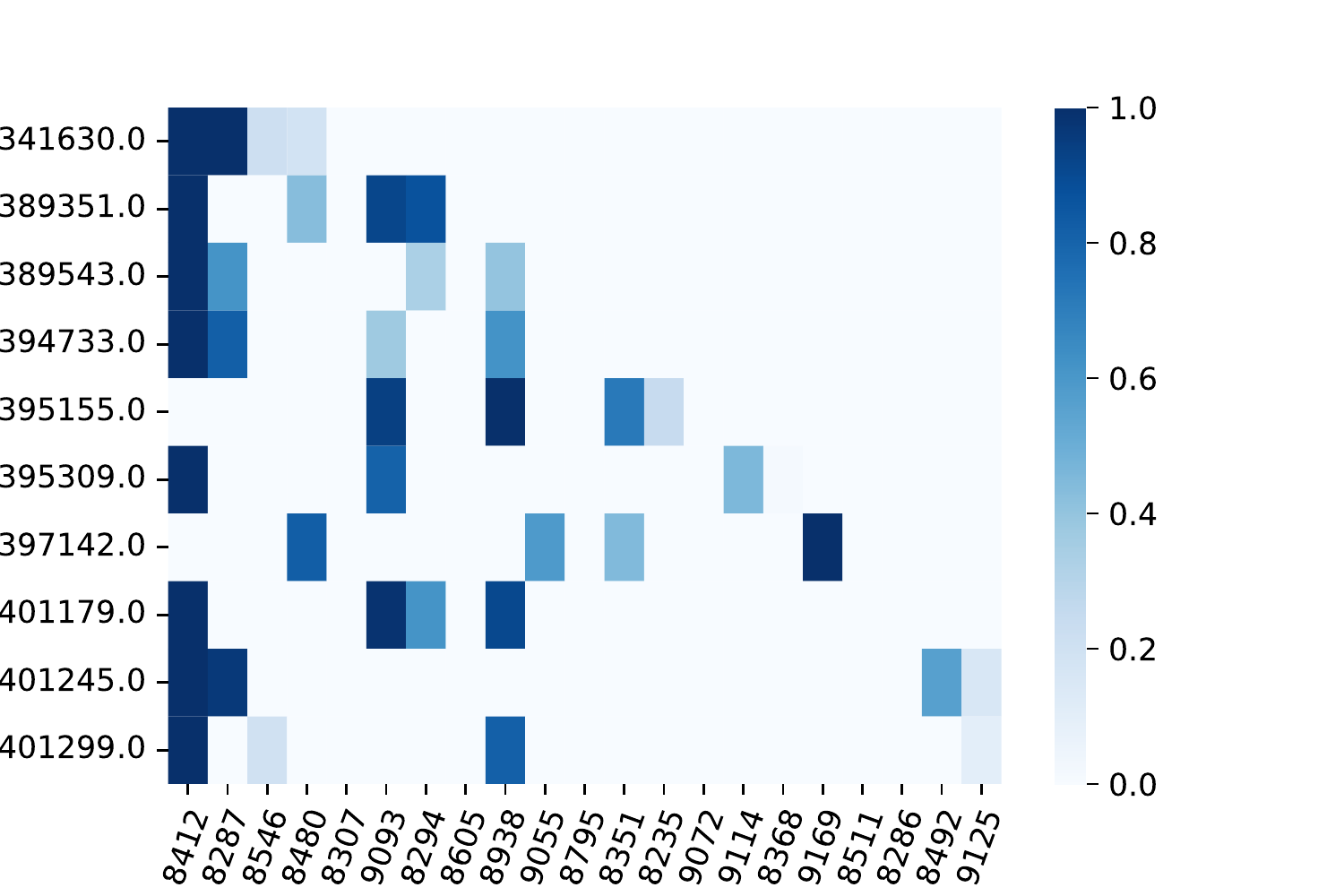}
}
\subfigure[Training set for user-416.]{
\includegraphics[width=0.47\linewidth]{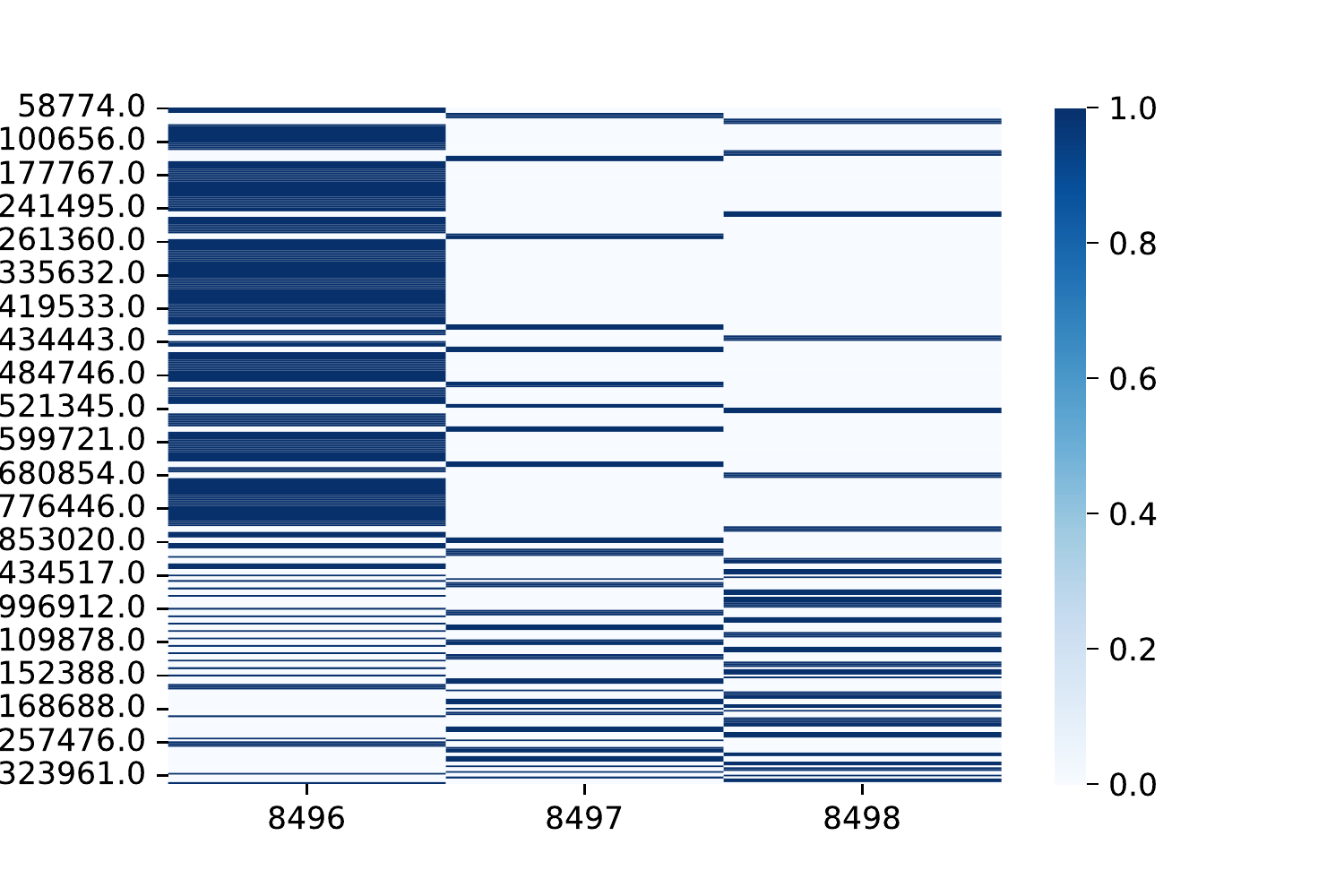}
}
\subfigure[Predictions for user-416.]{
\includegraphics[width=0.47\linewidth]{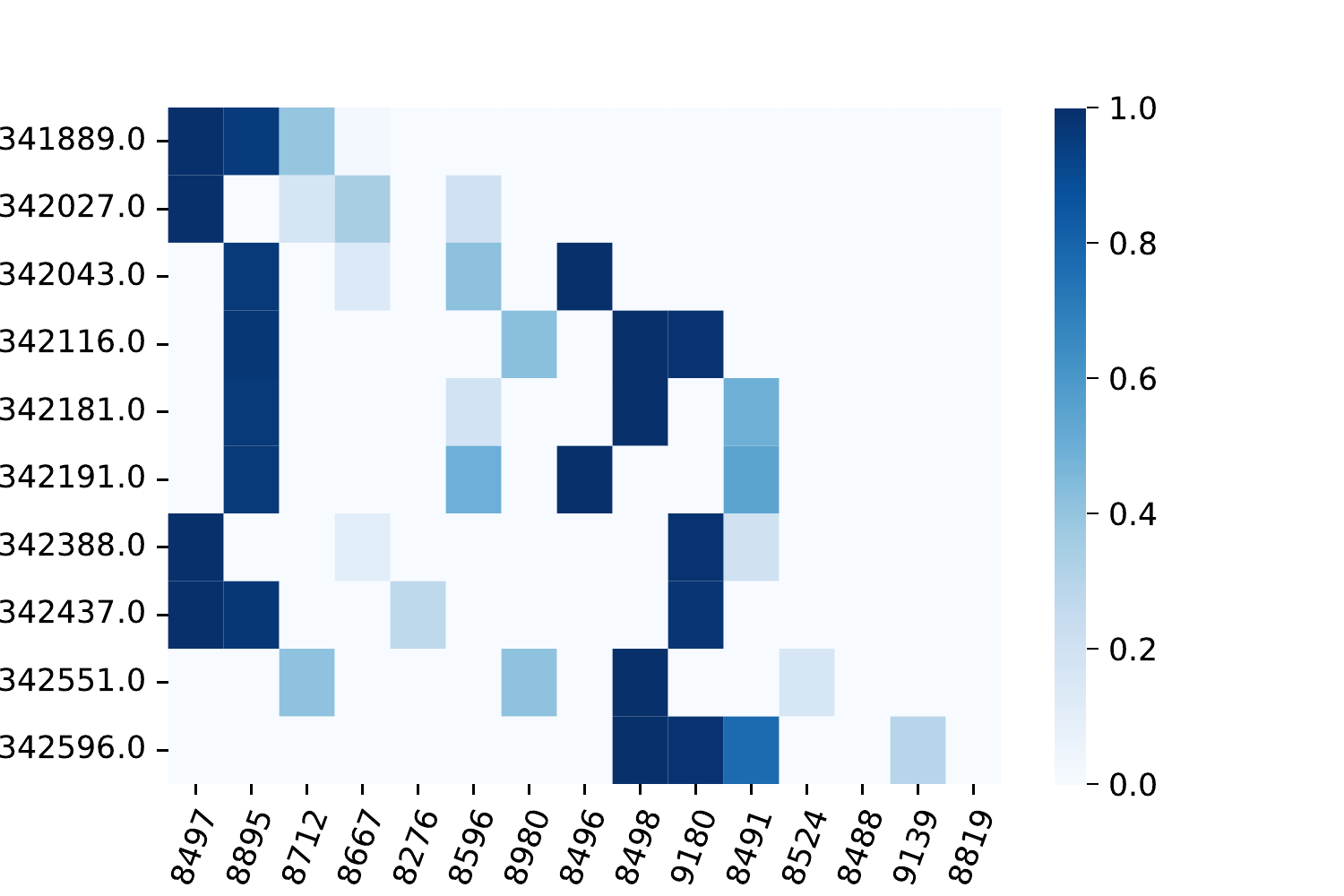}
}
\vspace{-0.15in}
\caption{Case study for specific users with visualization. The horizontal axis and vertical axis represent chronological timestamps and items respectively. (a) and (c) represent the training set for specific users, while (b) and (d) represent the prediction results of our \modelname.}
\label{fig:long short heatmap}
\vspace{-0.15in}
\end{figure}

\subsubsection{Case study for \modelname.}

In this part, we randomly selected two users (i.e. 320 and 416) from Wikipedia dataset, as well as their items they interacted. Figure \ref{fig:long short heatmap} shows the visualization of training set and predictions over timestamps by user and its relevant items. We find that: (1) there is a certain periodicity in the training set for users, i.e., users regularly interact with certain items. (2) in the test set, our \modelname captures this periodic pattern well, thanks to both long short-term preference modeling and memory mechanism.



\subsection{Ablation Study (RQ3)}\label{RQ3} 

We consider different model variants of \modelname from several perspectives and analyze their effects: (1) \modelname-no-short, which removes short-term embeddings and memory mechanism; (2) \modelname-sum, which replaces self-attention architecture with summation operation in DSACF; (3) \modelname-position, which replaces continuous-time encoding with position encoding; (4) \modelname-mean, which replaces the last time aggregation with mean aggregation in memory mechanism; (5) \modelname-2l, which sets the number of layers to 2.

We can observe that the full version of our developed \modelname achieves the best performance in all cases. As shown in Table \ref{tab:ablation}, we further summarize the conclusions: (1) memory mechanism significantly boost the performance of \modelname. For example, the performance \textit{w.r.t} Recall@10 improves from 0.7232 to 0.8635. (2) continuous-time embedding and DSACF plays a pivotal role in \modelname, capturing dynamic evolution information and collaborative signal. (3) one layer of our method is enough to capture long short-term preference, this is because the effect of more layers is not obvious.

\begin{table}[htp]
\vspace{-0.15in}
  \caption{Results of Ablation Study on Wikipedia. $\downarrow$ indicates a performance drop greater than 10\%.}
  \label{tab:ablation}
  \resizebox{\linewidth}{!}{
  \begin{tabular}{l|cc|cc|c}
    \toprule
    \multirow{2}{*}{Method} & \multicolumn{2}{c|}{Recall@K} & \multicolumn{2}{c|}{NDCG@K} & \multirow{2}{*}{MRR}  \\
    \cmidrule(lr){2-3}\cmidrule(lr){4-5}
    ~ & K=10 & K=20  & K=10 & K=20  & ~ \\
    \midrule
    \modelname-no-short & 0.7232$\downarrow$ & 0.7608$\downarrow$  & 0.6682$\downarrow$ & 0.6775$\downarrow$ &  0.6558$\downarrow$ \\
    \modelname-sum & 0.7824 & 0.8211 &  0.6920 & 0.7017 &  0.6686 \\
    \modelname-position & 0.8099 & 0.8458 &  0.7389 & 0.7479  & 0.7210 \\
    \modelname-mean & 0.8592 & 0.8833  & 0.8043 & 0.8103  & 0.7898 \\
    \midrule
   \textbf{\modelname} & \textbf{0.8635} & \textbf{0.8887}  & \textbf{0.8076} & \textbf{0.8139} & \textbf{0.7930} \\
   \textbf{\modelname-2l} & \textbf{0.8671} & \textbf{0.8893} &  \textbf{0.8197} & \textbf{0.8252}  & \textbf{0.8073} \\
  \bottomrule
\end{tabular}
}
\vspace{-0.15in}
\end{table}

\subsection{Efficiency of \modelname (RQ4)}\label{RQ4}
\begin{figure}[h]
\centering
\subfigure[Wikipedia dataset.]{
\includegraphics[width=0.46\linewidth]{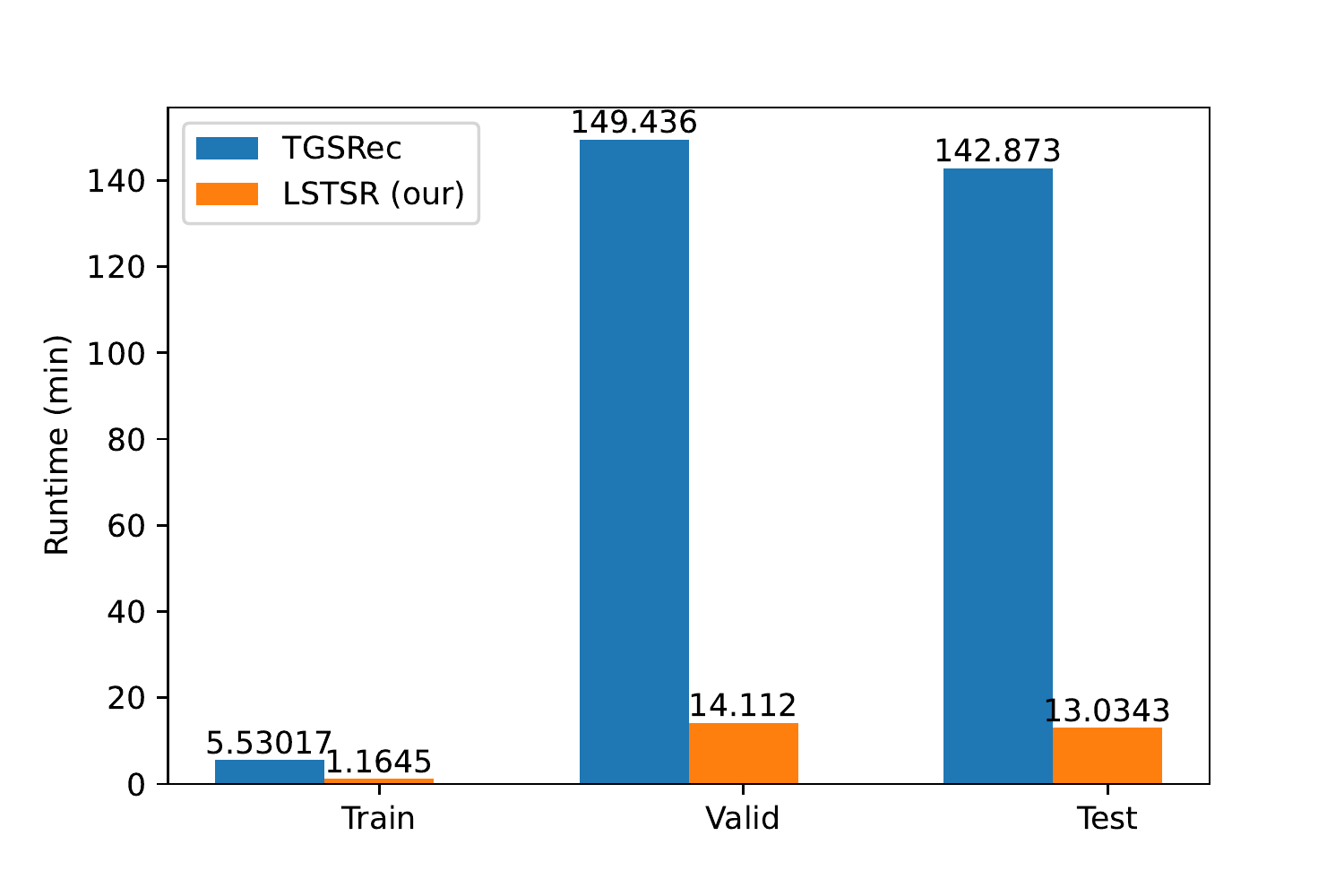}
}
\subfigure[Reddit dataset.]{
\includegraphics[width=0.46\linewidth]{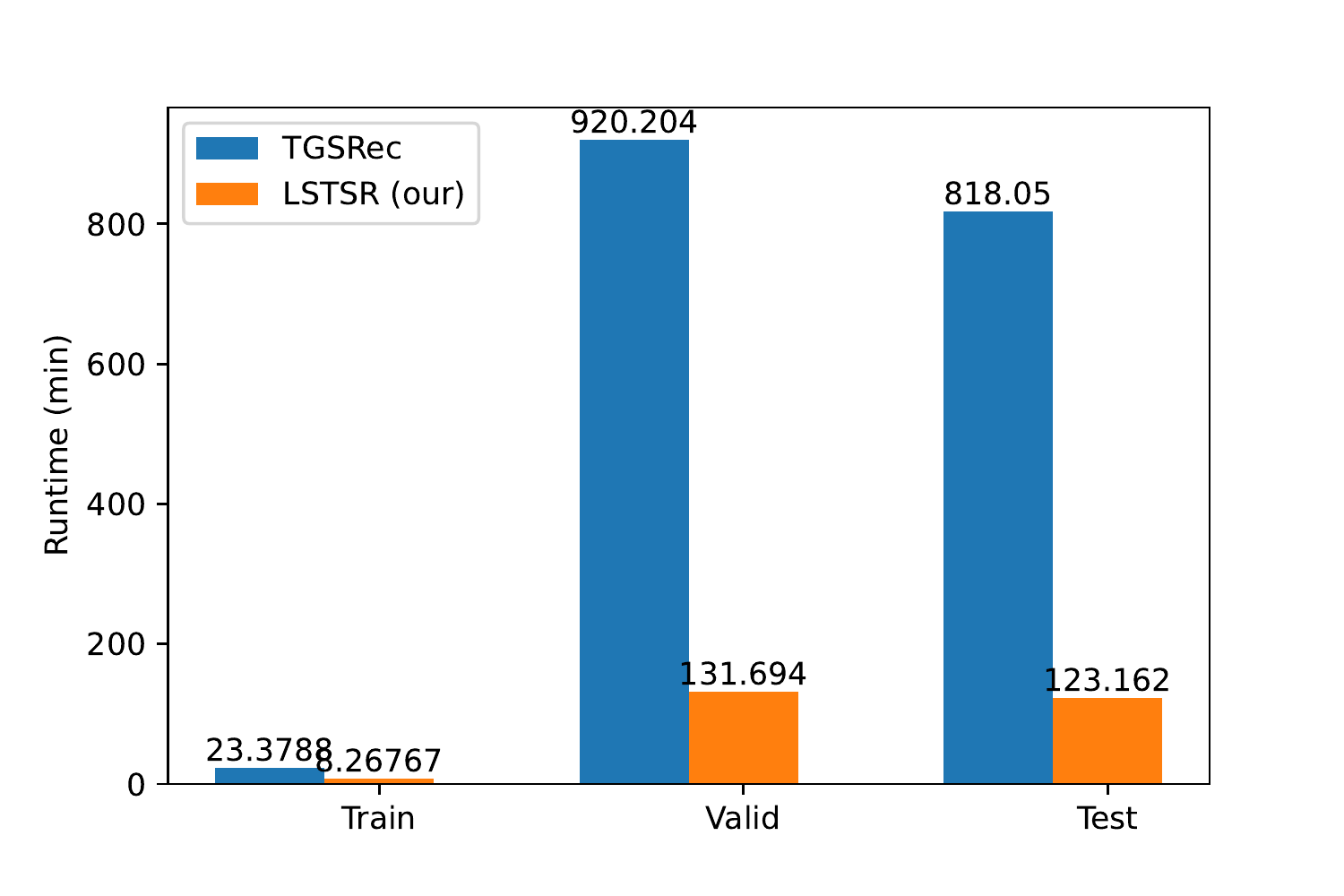}
}
\vspace{-0.15in}
\caption{Comparison with TGSRec and \modelname of the runtime (in minutes) per epoch (train, validation, and test) on Wikipedia and Reddit.}
\label{fig:time}
\vspace{-0.15in}
\end{figure}

In continuous-time sequential recommendation task, inference (validation/test) is the bottleneck of the performance, because graph-based representation of all candidate items is re-computed for everything at different timestamps. As shown in Figure \ref{fig:time}, in Wikipedia, training TGSRec requires 52x more time than inference, while that of our \modelname is 23x. In addition, similar observations can be drawn on Reddit. We compare the running time of \modelname with TGSRec, which focus on continuous-time sequential recommendation task. As shown in Figure \ref{fig:time}, \modelname can achieve near 5x and 10x speedup than TGSRec in training and inference, respectively. The reason is that we adopt memory mechanism, which represents the node’s history in a compressed format and maintains one-hop information. Therefore, one layer is enough. Furthermore, our dynamic neighbor sampling and last time aggregation also save a lot of time.

\section{Related Work}
\label{sec:related}


\subsection{Sequential Recommendation Methods} 
There exists some works for modeling fast-changing short-term preference in sequential recommendation, which could be roughly summarized into two categories.

One line of research attempts to capture short-term preference from user's last few interacted items. NARM \cite{li2017neural} and STAMP \cite{liu2018stamp} models user preference with two separate encoders, where the local encoder can select recent interacted items to capture short-term preference. SR-GNN \cite{wu2019sessionbased} also follows the same idea. SASRec \cite{kang2018selfattentive} and TiSASRec \cite{li2020time} adopts self-attention mechanism in history user behavior to adaptively capture short-term preference. 

Another line of research applies time series methods to capture the evolution of dynamic short-term preference from history user behavior. For example, GRU4Rec \cite{hidasi2016sessionbaseda}, DIEN \cite{zhou2019deep} and SLi-Rec \cite{yu2019adaptive} applies RNN \cite{medsker2001recurrent}, GRU \cite{chung2014empirical} and LSTM \cite{graves2012long}, respectively.


\subsection{Dynamic Graph-based Methods} 

Dynamic graph neural networks have been widely studied in many tasks, such as future interaction prediction and dynamic link prediction \cite{skarding2021foundations}. For example, in future interaction prediction, JODIE \cite{kumar2019predicting} uses RNN to capture user-item evolution; in dynamic link prediction, DyREP \cite{trivedi2019dyrep}, TGAT \cite{xu2020inductive} and TGN \cite{rossi2020temporal} propose variance dynamic networks. Recently, some dynamic graph-based methods for user preference modeling have been studied and achieved state-of-the-art performance by taking both graph structures and timestamps into consideration. TGSRec \cite{fan2021continuoustime} and DGSR \cite{zhang2022dynamic} unify sequential patterns and dynamic collaborative signals to capture the evolution of user-item interactions. These methods assume that user preference evolve smoothly and focus more on user's general/overall preference, which remains stable for a long time.

\section{Conclusion}
\label{sec:conclusion}

In this paper, we propose \modelname to capture the evolution of short-term preference for continuous-time sequential recommendation task. Based on existing dynamic graph-based methods, a novel memory mechanism is proposed to explicitly encode and update user's dynamic  short-term preference. Extensive experiment results show that our method captures the changing trend of short-term preference, and significantly outperforms various state-of-the-art 
recommendation models. Lastly, we will further study how to model user preference with long and noisy history behaviors better and streaming it for online industrial scenarios.

\section{Acknowledgments}
\label{sec:acks}

No acknowledgement. 

\bibliographystyle{aaai}
\bibliography{myref}

\begin{thebibliography}{}

\bibitem[\protect\citeauthoryear{An \bgroup et al\mbox.\egroup
  }{2019}]{an2019neural}
An, M.; Wu, F.; Wu, C.; Zhang, K.; Liu, Z.; and Xie, X.
\newblock 2019.
\newblock Neural news recommendation with long-and short-term user
  representations.
\newblock In {\em Proceedings of the 57th Annual Meeting of the Association for
  Computational Linguistics},  336--345.

\bibitem[\protect\citeauthoryear{Anderson, Butts, and
  Carley}{1999}]{anderson1999interaction}
Anderson, B.~S.; Butts, C.; and Carley, K.
\newblock 1999.
\newblock The interaction of size and density with graph-level indices.
\newblock {\em Social networks} 21(3):239--267.

\bibitem[\protect\citeauthoryear{Chang \bgroup et al\mbox.\egroup
  }{2021}]{chang2021sequential}
Chang, J.; Gao, C.; Zheng, Y.; Hui, Y.; Niu, Y.; Song, Y.; Jin, D.; and Li, Y.
\newblock 2021.
\newblock Sequential {{Recommendation}} with {{Graph Neural Networks}}.
\newblock In {\em Proceedings of the 44th {{International ACM SIGIR
  Conference}} on {{Research}} and {{Development}} in {{Information
  Retrieval}}},  378--387.
\newblock {Virtual Event Canada}: {ACM}.

\bibitem[\protect\citeauthoryear{Chen \bgroup et al\mbox.\egroup
  }{2018}]{chen2018sequential}
Chen, X.; Xu, H.; Zhang, Y.; Tang, J.; Cao, Y.; Qin, Z.; and Zha, H.
\newblock 2018.
\newblock Sequential recommendation with user memory networks.
\newblock In {\em Proceedings of the eleventh ACM international conference on
  web search and data mining},  108--116.

\bibitem[\protect\citeauthoryear{Chung \bgroup et al\mbox.\egroup
  }{2014}]{chung2014empirical}
Chung, J.; Gulcehre, C.; Cho, K.; and Bengio, Y.
\newblock 2014.
\newblock Empirical evaluation of gated recurrent neural networks on sequence
  modeling.
\newblock {\em arXiv preprint arXiv:1412.3555}.

\bibitem[\protect\citeauthoryear{Fan \bgroup et al\mbox.\egroup
  }{2021}]{fan2021continuoustime}
Fan, Z.; Liu, Z.; Zhang, J.; Xiong, Y.; Zheng, L.; and Yu, P.~S.
\newblock 2021.
\newblock Continuous-{{Time Sequential Recommendation}} with {{Temporal Graph
  Collaborative Transformer}}.
\newblock In {\em Proceedings of the 30th {{ACM International Conference}} on
  {{Information}} \& {{Knowledge Management}}}. {New York, NY, USA}:
  {Association for Computing Machinery}.
\newblock  433--442.

\bibitem[\protect\citeauthoryear{Glorot and
  Bengio}{2010}]{glorot2010understanding}
Glorot, X., and Bengio, Y.
\newblock 2010.
\newblock Understanding the difficulty of training deep feedforward neural
  networks.
\newblock In {\em Proceedings of the thirteenth international conference on
  artificial intelligence and statistics},  249--256.
\newblock JMLR Workshop and Conference Proceedings.

\bibitem[\protect\citeauthoryear{Graves}{2012}]{graves2012long}
Graves, A.
\newblock 2012.
\newblock Long short-term memory.
\newblock {\em Supervised sequence labelling with recurrent neural networks}
  37--45.

\bibitem[\protect\citeauthoryear{He \bgroup et al\mbox.\egroup
  }{2017}]{he2017neural}
He, X.; Liao, L.; Zhang, H.; Nie, L.; Hu, X.; and Chua, T.-S.
\newblock 2017.
\newblock Neural collaborative filtering.
\newblock In {\em Proceedings of the 26th international conference on world
  wide web},  173--182.

\bibitem[\protect\citeauthoryear{He \bgroup et al\mbox.\egroup
  }{2020}]{he2020lightgcn}
He, X.; Deng, K.; Wang, X.; Li, Y.; Zhang, Y.; and Wang, M.
\newblock 2020.
\newblock Lightgcn: {{Simplifying}} and powering graph convolution network for
  recommendation.
\newblock In {\em Proceedings of the 43rd {{International ACM SIGIR}}
  Conference on Research and Development in {{Information Retrieval}}},
  639--648.

\bibitem[\protect\citeauthoryear{Hidasi \bgroup et al\mbox.\egroup
  }{2016}]{hidasi2016sessionbaseda}
Hidasi, B.; Karatzoglou, A.; Baltrunas, L.; and Tikk, D.
\newblock 2016.
\newblock Session-based {{Recommendations}} with {{Recurrent Neural Networks}}.
\newblock In Bengio, Y., and LeCun, Y., eds., {\em 4th {{International
  Conference}} on {{Learning Representations}}, {{ICLR}} 2016, {{San Juan}},
  {{Puerto Rico}}, {{May}} 2-4, 2016, {{Conference Track Proceedings}}}.

\bibitem[\protect\citeauthoryear{Horn}{1990}]{horn1990hadamard}
Horn, R.~A.
\newblock 1990.
\newblock The hadamard product.
\newblock In {\em Proc. Symp. Appl. Math}, volume~40,  87--169.

\bibitem[\protect\citeauthoryear{Hsieh \bgroup et al\mbox.\egroup
  }{2017}]{hsieh2017collaborative}
Hsieh, C.-K.; Yang, L.; Cui, Y.; Lin, T.-Y.; Belongie, S.; and Estrin, D.
\newblock 2017.
\newblock Collaborative metric learning.
\newblock In {\em Proceedings of the 26th international conference on world
  wide web},  193--201.

\bibitem[\protect\citeauthoryear{Kang and
  McAuley}{2018}]{kang2018selfattentive}
Kang, W.-C., and McAuley, J.
\newblock 2018.
\newblock Self-{{Attentive Sequential Recommendation}}.
\newblock In {\em 2018 {{IEEE International Conference}} on {{Data Mining}}
  ({{ICDM}})},  197--206.

\bibitem[\protect\citeauthoryear{Kumar, Zhang, and
  Leskovec}{2019}]{kumar2019predicting}
Kumar, S.; Zhang, X.; and Leskovec, J.
\newblock 2019.
\newblock Predicting {{Dynamic Embedding Trajectory}} in {{Temporal Interaction
  Networks}}.
\newblock In {\em Proceedings of the 25th {{ACM SIGKDD International
  Conference}} on {{Knowledge Discovery}} \& {{Data Mining}}},  1269--1278.
\newblock {Anchorage AK USA}: {ACM}.

\bibitem[\protect\citeauthoryear{Li \bgroup et al\mbox.\egroup
  }{2017}]{li2017neural}
Li, J.; Ren, P.; Chen, Z.; Ren, Z.; Lian, T.; and Ma, J.
\newblock 2017.
\newblock Neural attentive session-based recommendation.
\newblock In {\em Proceedings of the 2017 {{ACM}} on {{Conference}} on
  {{Information}} and {{Knowledge Management}}},  1419--1428.

\bibitem[\protect\citeauthoryear{Li, Wang, and McAuley}{2020}]{li2020time}
Li, J.; Wang, Y.; and McAuley, J.
\newblock 2020.
\newblock Time {{Interval Aware Self-Attention}} for {{Sequential
  Recommendation}}.
\newblock In {\em Proceedings of the 13th {{International Conference}} on {{Web
  Search}} and {{Data Mining}}},  322--330.
\newblock {Houston TX USA}: {ACM}.

\bibitem[\protect\citeauthoryear{Liu \bgroup et al\mbox.\egroup
  }{2018}]{liu2018stamp}
Liu, Q.; Zeng, Y.; Mokhosi, R.; and Zhang, H.
\newblock 2018.
\newblock {{STAMP}}: {{Short-Term Attention}}/{{Memory Priority Model}} for
  {{Session-based Recommendation}}.
\newblock In {\em Proceedings of the 24th {{ACM SIGKDD International
  Conference}} on {{Knowledge Discovery}} \& {{Data Mining}}},  1831--1839.
\newblock {London United Kingdom}: {ACM}.

\bibitem[\protect\citeauthoryear{Loomis}{2013}]{loomis2013introduction}
Loomis, L.~H.
\newblock 2013.
\newblock {\em Introduction to abstract harmonic analysis}.
\newblock Courier Corporation.

\bibitem[\protect\citeauthoryear{McAuley \bgroup et al\mbox.\egroup
  }{2015}]{mcauley2015image}
McAuley, J.; Targett, C.; Shi, Q.; and Van Den~Hengel, A.
\newblock 2015.
\newblock Image-based recommendations on styles and substitutes.
\newblock In {\em Proceedings of the 38th international ACM SIGIR conference on
  research and development in information retrieval},  43--52.

\bibitem[\protect\citeauthoryear{Medsker and Jain}{2001}]{medsker2001recurrent}
Medsker, L.~R., and Jain, L.
\newblock 2001.
\newblock Recurrent neural networks.
\newblock {\em Design and Applications} 5:64--67.

\bibitem[\protect\citeauthoryear{Paszke \bgroup et al\mbox.\egroup
  }{2019}]{paszke2019pytorch}
Paszke, A.; Gross, S.; Massa, F.; Lerer, A.; Bradbury, J.; Chanan, G.; Killeen,
  T.; Lin, Z.; Gimelshein, N.; Antiga, L.; et~al.
\newblock 2019.
\newblock Pytorch: An imperative style, high-performance deep learning library.
\newblock {\em Advances in neural information processing systems} 32.

\bibitem[\protect\citeauthoryear{Pi \bgroup et al\mbox.\egroup
  }{2019}]{pi2019practice}
Pi, Q.; Bian, W.; Zhou, G.; Zhu, X.; and Gai, K.
\newblock 2019.
\newblock Practice on long sequential user behavior modeling for click-through
  rate prediction.
\newblock In {\em Proceedings of the 25th ACM SIGKDD International Conference
  on Knowledge Discovery \& Data Mining},  2671--2679.

\bibitem[\protect\citeauthoryear{Rendle \bgroup et al\mbox.\egroup
  }{2012}]{rendle2012bpr}
Rendle, S.; Freudenthaler, C.; Gantner, Z.; and Schmidt-Thieme, L.
\newblock 2012.
\newblock Bpr: Bayesian personalized ranking from implicit feedback.
\newblock {\em arXiv preprint arXiv:1205.2618}.

\bibitem[\protect\citeauthoryear{Rossi \bgroup et al\mbox.\egroup
  }{2020}]{rossi2020temporal}
Rossi, E.; Chamberlain, B.; Frasca, F.; Eynard, D.; Monti, F.; and Bronstein,
  M.
\newblock 2020.
\newblock Temporal {{Graph Networks}} for {{Deep Learning}} on {{Dynamic
  Graphs}}.
\newblock {\em arXiv:2006.10637 [cs, stat]}.

\bibitem[\protect\citeauthoryear{Sazli}{2006}]{sazli2006brief}
Sazli, M.~H.
\newblock 2006.
\newblock A brief review of feed-forward neural networks.
\newblock {\em Communications Faculty of Sciences University of Ankara Series
  A2-A3 Physical Sciences and Engineering} 50(01).

\bibitem[\protect\citeauthoryear{Skarding, Gabrys, and
  Musial}{2021}]{skarding2021foundations}
Skarding, J.; Gabrys, B.; and Musial, K.
\newblock 2021.
\newblock Foundations and {{Modeling}} of {{Dynamic Networks Using Dynamic
  Graph Neural Networks}}: {{A Survey}}.
\newblock {\em IEEE Access} 9:79143--79168.

\bibitem[\protect\citeauthoryear{Trivedi \bgroup et al\mbox.\egroup
  }{2019}]{trivedi2019dyrep}
Trivedi, R.; Farajtabar, M.; Biswal, P.; and Zha, H.
\newblock 2019.
\newblock Dyrep: {{Learning}} representations over dynamic graphs.
\newblock In {\em International Conference on Learning Representations}.

\bibitem[\protect\citeauthoryear{Vaswani \bgroup et al\mbox.\egroup
  }{2017}]{vaswani2017attention}
Vaswani, A.; Shazeer, N.; Parmar, N.; Uszkoreit, J.; Jones, L.; Gomez, A.~N.;
  Kaiser, {\L}.; and Polosukhin, I.
\newblock 2017.
\newblock Attention is all you need.
\newblock {\em Advances in neural information processing systems} 30.

\bibitem[\protect\citeauthoryear{Wang \bgroup et al\mbox.\egroup
  }{2019}]{wang2019neural}
Wang, X.; He, X.; Wang, M.; Feng, F.; and Chua, T.-S.
\newblock 2019.
\newblock Neural {{Graph Collaborative Filtering}}.
\newblock {\em Proceedings of the 42nd International ACM SIGIR Conference on
  Research and Development in Information Retrieval}  165--174.

\bibitem[\protect\citeauthoryear{Wu \bgroup et al\mbox.\egroup
  }{2019}]{wu2019sessionbased}
Wu, S.; Tang, Y.; Zhu, Y.; Wang, L.; Xie, X.; and Tan, T.
\newblock 2019.
\newblock Session-based {{Recommendation}} with {{Graph Neural Networks}}.
\newblock {\em AAAI} 33:346--353.

\bibitem[\protect\citeauthoryear{Xu \bgroup et al\mbox.\egroup
  }{2020}]{xu2020inductive}
Xu, D.; Ruan, C.; Korpeoglu, E.; Kumar, S.; and Achan, K.
\newblock 2020.
\newblock Inductive representation learning on temporal graphs.
\newblock In {\em International {{Conference}} on {{Learning
  Representations}}}.

\bibitem[\protect\citeauthoryear{Yu \bgroup et al\mbox.\egroup
  }{2019}]{yu2019adaptive}
Yu, Z.; Lian, J.; Mahmoody, A.; Liu, G.; and Xie, X.
\newblock 2019.
\newblock Adaptive {{User Modeling}} with {{Long}} and {{Short-Term
  Preferences}} for {{Personalized Recommendation}}.
\newblock In {\em Proceedings of the {{Twenty-Eighth International Joint
  Conference}} on {{Artificial Intelligence}}},  4213--4219.
\newblock {Macao, China}: {International Joint Conferences on Artificial
  Intelligence Organization}.

\bibitem[\protect\citeauthoryear{Zhang \bgroup et al\mbox.\egroup
  }{2022}]{zhang2022dynamic}
Zhang, M.; Wu, S.; Yu, X.; Liu, Q.; and Wang, L.
\newblock 2022.
\newblock Dynamic graph neural networks for sequential recommendation.
\newblock {\em IEEE Transactions on Knowledge and Data Engineering}.

\bibitem[\protect\citeauthoryear{Zheng \bgroup et al\mbox.\egroup
  }{2022}]{zheng2022disentangling}
Zheng, Y.; Gao, C.; Chang, J.; Niu, Y.; Song, Y.; Jin, D.; and Li, Y.
\newblock 2022.
\newblock Disentangling long and short-term interests for recommendation.
\newblock In {\em Proceedings of the ACM Web Conference 2022},  2256--2267.

\bibitem[\protect\citeauthoryear{Zhou \bgroup et al\mbox.\egroup
  }{2018}]{zhou2018deep}
Zhou, G.; Zhu, X.; Song, C.; Fan, Y.; Zhu, H.; Ma, X.; Yan, Y.; Jin, J.; Li,
  H.; and Gai, K.
\newblock 2018.
\newblock Deep interest network for click-through rate prediction.
\newblock In {\em Proceedings of the 24th ACM SIGKDD international conference
  on knowledge discovery \& data mining},  1059--1068.

\bibitem[\protect\citeauthoryear{Zhou \bgroup et al\mbox.\egroup
  }{2019}]{zhou2019deep}
Zhou, G.; Mou, N.; Fan, Y.; Pi, Q.; Bian, W.; Zhou, C.; Zhu, X.; and Gai, K.
\newblock 2019.
\newblock Deep {{Interest Evolution Network}} for {{Click-Through Rate
  Prediction}}.
\newblock {\em AAAI} 33:5941--5948.

\end{thebibliography}

\appendix


\end{document}